\documentclass[11pt]{amsart} 
\usepackage{graphicx}
\begin{document}

\title{Construction of an isotropic cellular automaton for a reaction-diffusion equation by means of a random walk}

\author{A. NISHIYAMA and T. TOKIHIRO}

\address{Graduate School of Mathematical Sciences, The University of Tokyo, 3-8-1 Komaba, Meguro-ku, Tokyo 153-8914, Japan}

\begin{abstract}
We propose a new method to construct an isotropic cellular automaton corresponding to a reaction-diffusion equation.
The method consists of replacing the diffusion term and  the reaction term of the reaction-diffusion equation with a random walk of microscopic particles and a discrete vector field which defines the time evolution of the particles.
The cellular automaton thus obtained can retain isotropy and therefore reproduces the patterns found in the numerical solutions of the reaction-diffusion equation.
As a specific example, we apply the method to the Belousov-Zhabotinsky reaction in excitable media.

\end{abstract}
\maketitle

\section{introduction}
Reaction-diffusion equations are used extensively for modeling pattern formation observed in natural and social phenomena.
The equations are deduced from the simple idea that concentration changes of a material in a system are caused by reactions between the materials contained in the system and by diffusion of each material.
Numerical solutions of the equations show a variety of patterns, and they are used widely in various fields such as chemistry, biology, medical science etc. \cite{murray}.

An alternative approach to modeling pattern formation is to use a cellular automaton (hereafter abbreviated as CA) \cite{wolfram}.
CAs are mathematical models with discrete time, space and state variables, and are defined by simple time evolution rules.
They are able to reproduce complex patterns, and therefore have a lot of applications.
For example, the lattice gas CAs represent very well the various features of fluid dynamics and reaction-diffusion phenomena \cite{latticegas}.
In general, CAs have the advantage of being able to simulate phenomena at lower computational cost than when using partial differential equations.

A lot of models for cooperative phenomena of excitable media, derived by using partial differential equations or cellular automata have been proposed.
The so-called "oregonator" model  \cite{field, tyson2, jahnke}  expressed in terms of differential equations can explain the Belousov-Zhabotinsky(BZ) reaction \cite{zhabotinsky} known as a typical example of a reaction-diffusion system. On the other hand, some approaches by means of cellular automata for excitable media have been proposed in \cite{moe, greenberg, greenberg2}.
However, these early CA models featured update rules based on nearest neighbor connections, and hence, faced several serious shortcomings \cite{winfree}. 
The most serious of these is the lack of curvature and dispersion effects and unwanted anisotropy of the front motion \cite{gerhardt, markus}.
To overcome these problems, several automata models have been proposed \cite{gerhardt, gerhardt2, gerhardt3, markus, weimar, weimar2}.
Gerhardt, Shuster and Tyson introduced bigger neighborhoods to model the curvature effects and make the threshold a linear function to take dispersion into account \cite{gerhardt, gerhardt2, gerhardt3}.
Weimar, Tyson and Watson improved the models by introducing a mask, i.e. a weighted summation of automaton values over large neighborhoods \cite{weimar, weimar2}, and Henze and Tyson extended it to three spatial dimensions \cite{henze}.
These models recover curvature and dispersion effects well, but the anisotropy of wave propagation is not completely eliminated.

In cellular automata, to recover the isotropy of the time-evolution patterns is in fact a difficult problem.
If the adopted lattice used in the modeling has periodicity, such as a square lattice or a hexagonal lattice, then, due to this periodicity the time-evolution patterns obtained from the model become anisotropic.
In \cite{markus2} Markus and Hess have proposed an isotropic model for excitable media.
In their model, they use a square lattice, but instead of placing each grid point at the center of a unit cell, they assign each grid point to a random location within its unit cell.
The isotropy is recovered by taking a large number of neighboring cells within a circular area with radius R.
As other CA models for excitable media which seem to recover anisotropy to some extent, one can cite the "Moving Average CA" method by Weimar \cite{weimar3}, the lattice gas method CAs by Raymond et al. \cite{raymond} or Chen \cite{chen} et al., the isotropic CA model for the growth process of a bacterial colony by using a Voronoi lattice \cite{voronoi} proposed in \cite{badoual}.
In the previous paper \cite{nishiyama}, Tanaka and the authors also considered an isotropic CA model for the BZ reaction.
This simple square lattice model adopts a Moore neighborhood with radius 1 as for the neighborhood cells and recovers the isotropy by introducing spatial randomness in relation to a threshold controlling the excitation of reaction.

In the present paper, we propose a new method to construct a CA model corresponding to a reaction-diffusion system. The method is based on the random walk and on the phase diagram of reaction equations. 
We require the method to satisfy the following two conditions:
(i)The time evolution pattern of the CA preserves the isotropy of the reaction-diffusion equation.
(ii)The CA model explicitly contains the control parameters corresponding to the reaction terms and diffusion terms in the reaction-diffusion equation.

In the next section, we explain the general method to construct the CA corresponding to a given reaction-diffusion equation.
In section 3, the method introduced in section 2 is applied to the BZ reaction, and the obtained evolution patterns are examined.
A summary or the results is given in section 4.

\section{methodology}
In this section, a method to construct the CA corresponding to a given reaction-diffusion equation is introduced.
The idea is very simple: to replace the diffusion term by a random walk process and the reaction terms by time evolution along a discrete vector field obtained from the phase diagram.
First, we explain the reaction-diffusion equation in brief, then we introduce our CA model.

\subsection{Reaction-Diffusion Equation}
Suppose that there are $N$ different reactive materials $U_{1}$, $U_{2}$, $\cdots$, $U_{N}$ in some spatial region.
Let the densities of these materials at position $\boldsymbol{r}$ and time $t$ be $u_{1}(\boldsymbol{r},t)$, $u_{2}(\boldsymbol{r},t)$, $\cdots$, $u_{N}(\boldsymbol{r},t)$ respectively, and put $\boldsymbol{u}:=$$(u_{1}$, $u_{2}$, $\cdots$, $u_{N})^{T}$ $\in$ $\Bbb{R}^{N}$.
The reaction-diffusion equation is given as
\begin{equation}\label{rde}
\frac{\partial \boldsymbol{u}}{\partial t} =\boldsymbol{f}(\boldsymbol{u})+D \nabla^{2} \boldsymbol{u},
\end{equation}
where $D={\rm{diag}}(d_{1}, d_{2}, \cdots, d_{N})$ i.e. $D$ is a $N \times N$ diagonal matrix which has the diffusion coefficient $d_{j}$ for each material as a diagonal element. 
The vector $\boldsymbol{f}(\boldsymbol{u})=$ $(f_{1}(\boldsymbol{u})$, $f_{2}(\boldsymbol{u})$, $\cdots$, 
$f_{N}(\boldsymbol{u}))^{T}$ expresses the interactions between the materials 
and $\nabla:=\frac{\partial}{\partial \boldsymbol{r}}$ is the nabla symbol, 
in particular $\nabla^{2}=\frac{\partial^{2}}{{\partial x}^{2}} + \frac{\partial^{2}}{{\partial y}^{2}}$ in two spatial dimension: $\boldsymbol{r}=(x,y)$.

\subsection{The CA Model}
We now introduce our cellular automaton model for the reaction-diffusion equation (\ref{rde}).
For simplicity, we assume that the equation is defined in two spatial dimensions and that the CA model is defined on a
two dimensional square lattice.
Generalization to higher dimensional systems and different types of lattices will be straightforward. 
In the reaction-diffusion equation (\ref{rde}), the variable $\boldsymbol{u}$ represents the density of reactive material.
In our CA model, we replace the density of material $\boldsymbol{u}(\boldsymbol{r},t)$ by the number of microscopic particles $\boldsymbol{u}^{t}_{mn} \in \Bbb{Z_{+}}^{N}$ at the corresponding lattice point $(m,n) \in \Bbb{Z}^2$ at time step $t \in \Bbb{Z}/2$.
Then the time evolution of our CA model is determined by
\begin{equation}
\begin{split}
&\boldsymbol{u}^{t+1/2}_{mn} = \boldsymbol{u}^{t}_{mn} +   \boldsymbol{R}(\boldsymbol{\hat{u}}^{t}_{mn}),\\
&\boldsymbol{u}^{t+1}_{mn} = \boldsymbol{u}^{t+1/2}_{mn} + \boldsymbol{F}(\boldsymbol{u}^{t+1/2}_{mn}),
\end{split}
\end{equation}
where $\boldsymbol{\hat{u}}^{t}_{mn}$ denotes a set of concentration variables around ($m,n$) at time step $t$, $\boldsymbol{R}(\boldsymbol{\hat{u}}^{t}_{mn})$ and $\boldsymbol{F}(\boldsymbol{u}^{t}_{mn})$ denote discrete vector fields which are obtained by replacing the diffusion and reaction terms of the reaction-diffusion equation respectively with the following processes:

\textbf{(i)}
The diffusion term $D \nabla^{2} \boldsymbol{u} $ corresponds to a random walk of particles.
Let us consider a random walk of particles defined in the Neumann neighborhood $\boldsymbol{\hat{u}}^{t}_{mn}=$ $\{\boldsymbol{u}^{t}_{m,n},$ $ \boldsymbol{u}^{t}_{m-1,n},$ $ \boldsymbol{u}^{t}_{m+1,n},$ $ \boldsymbol{u}^{t}_{m,n-1},$ $ \boldsymbol{u}^{t}_{m,n+1}\}$. One can equally adopt the Moor neighborhood or other neighborhoods.
We denote by $\boldsymbol{U}^{\rightarrow t}_{mn}$ the stochastic variable which defines the number of particles   moving from the $(m,n)$ site to the $(m+1,n)$ site due to the random walk of the particles, by $\boldsymbol{U}^{\rightarrow t}_{m-1,n}$ that from from $(m-1,n)$ to $(m,n)$, and so on.
Then $\boldsymbol{R}(\boldsymbol{\hat{u}}^{t}_{mn})$ equals the difference between the number of outgoing particles and that of the incoming particles due to the random walk at position $(m,n)$ and time $t$:
\begin{eqnarray}
\boldsymbol{R}(\boldsymbol{\hat{u}}^{t}_{mn}) = \boldsymbol{U}^{\rightarrow t}_{m-1,n} + \boldsymbol{U}^{\leftarrow t}_{m+1,n} + \boldsymbol{U}^{\uparrow t}_{m,n-1} + \boldsymbol{U}^{\downarrow t}_{m,n+1}\nonumber\\
- (\boldsymbol{U}^{\rightarrow t}_{mn} + \boldsymbol{U}^{\leftarrow t}_{mn} + \boldsymbol{U}^{\uparrow t}_{mn} + \boldsymbol{U}^{\downarrow t}_{mn}).
\end{eqnarray}
If we define $p_{j}$ as the transition probability of particles to one neighboring cell in the random walk, $1-4p_{j}$ is the probability of particles to stay on site and the expectation of $\boldsymbol{U}^{\rightarrow t}_{mn}$, $\boldsymbol{U}^{\rightarrow t}_{m-1,n}$ etc. are given respectively by $\langle \boldsymbol{U}^{\rightarrow t}_{m,n} \rangle=P\boldsymbol{u}^{t}_{m-1,n}$, $\langle \boldsymbol{U}^{\rightarrow t}_{m-1,n} \rangle=P\boldsymbol{u}^{t}_{m-1,n}$ with a diagonal matrix $P={\rm{diag}}(p_{1}, p_{2}, \cdots, p_{N})$.
We can see that $P$ corresponds to the diffusion coefficient $D$ of the reaction-diffusion equation.

\textbf{(ii)}
The reaction term $\boldsymbol{f}(\boldsymbol{u})$ is replaced with an appropriate discrete function $\boldsymbol{F}(\boldsymbol{u}^{t}_{mn})=$ $(F_{1}(\boldsymbol{u}^{t}_{mn})$, $F_{2}(\boldsymbol{u}^{t}_{mn})$, $\cdots$, $F_{N}(\boldsymbol{u}^{t}_{mn}))^{T}$ $\in$ $\Bbb{Z}^{N}$.
In equation (\ref{rde}), the reaction term $\boldsymbol{f}(\boldsymbol{u})$ is the vector field that defines the velocity vector $\frac{\partial \boldsymbol{u}}{\partial t}$.
Hence $\boldsymbol{F}(\boldsymbol{x})$ should be so chosen such that the time evolution of $\boldsymbol{x}$ is consistent with the typical orbits in the phase diagram of the ordinary differential equation
\[
\frac{d \boldsymbol{u}}{d t} =\boldsymbol{f}(\boldsymbol{u}).
\]

Since $\boldsymbol{F}(\boldsymbol{u}^{t}_{mn})$ sometimes returns a negative number, it is practically convenient to use the discrete vector field, $\boldsymbol{G}(\boldsymbol{u}^{t}_{mn}) :=\boldsymbol{u}^{t}_{mn} + \boldsymbol{F}(\boldsymbol{u}^{t}_{mn}) \geq 0$.
Then our CA model is rewritten as
\begin{equation}
\begin{split}
\label{camodel}
&\boldsymbol{u}^{t+1/2}_{mn} = \boldsymbol{u}^{t}_{mn} +   \boldsymbol{R}(\boldsymbol{\hat{u}}^{t}_{mn}), \\
&\boldsymbol{u}^{t+1}_{mn} = \boldsymbol{G}(\boldsymbol{u}^{t+1/2}_{mn}).
\end{split}
\end{equation}
We can obtain the time evolution pattern by successive substitution of  $\boldsymbol{R}(\boldsymbol{\hat{u}}^{t}_{mn})$ and $\boldsymbol{G}(\boldsymbol{u}^{t}_{mn})$ for appropriate initial conditions.

It is well known that the continuous limit of the random walk is equivalent to a diffusion equation,
and that in the large scale limit, isotropy of the distribution function of the particles is guaranteed by a random walk.
Since the discrete vector field $\boldsymbol{F}(\boldsymbol{u})$ is chosen such that it is essentially equivalent to the vector field $\boldsymbol{f}(\boldsymbol{u})$ in the continuous limit, we expect that the patterns obtained from our CA model become almost isotropic in certain large systems.
It should be noted that our model naturally contains the parameters $\{p_i\}$ which correspond to the diffusion coefficients and all other control parameters for reactions, which are necessarily contained in the discrete vector field.  

\section{CA model for the Belousov-Zhabotinsky reaction}
In this section, we apply the method introduced in the previous section to the Belousov-Zhabotinsky (BZ) reaction as a specific example.
First, we briefly explain the BZ reaction and oregonator known as a mathematical model for this reaction. Next, we introduce our CA model of the BZ reaction.

\subsection{BZ Reaction}
The BZ reaction is known as an oscillating oxidation-reduction reaction which occurs by mixing some chemical compounds (such as $\mathrm{Ce^{4+}}$, $\mathrm{BrO^{3-}}$, $\mathrm{CH_{2}(COOH)_{2}}$, $\mathrm{H_{2}SO_{4}}$).
If the BZ reaction is spread spatially, then it forms trigger waves, spiral waves or target patterns \cite{zhabotinsky}.
The BZ reaction is often modeled by using partial differential equations.
Among them, the oregonator, which is a system of simultaneous ordinary differential equations with two variables, is widely considered to be the simplest possible model \cite{tyson2}.
The time evolution of the spatial patterns in the BZ reaction is described by the following equation which adds diffusion terms to the oregonator:
\begin{equation}
\begin{split}
\label{oregonator}
\frac{\partial u}{\partial t} &= \frac{1}{\epsilon} \bigg[ u(1-u)-\frac{b v(u-a)}{u+a} \bigg] + d_{u} \nabla^{2} u,\\
\frac{\partial v}{\partial t} &= u-v + d_{v} \nabla^{2} v,
\end{split}
\end{equation}
where $b$ and $\epsilon$ (or $1/\epsilon$) are, respectively, a threshold which gives the excitation and a parameter which defines the excitability of reaction respectively. Depending on the parameters $a$, $b$ and $\epsilon$, Eq. (\ref{oregonator}) shows two typical states: an excitable state with one stable equilibrium point and an oscillatory state with one unstable equilibrium point.
Here we consider only the excitable state (or excitable media).

\begin{figure}[h]
\begin{center}
\includegraphics[width=8cm]{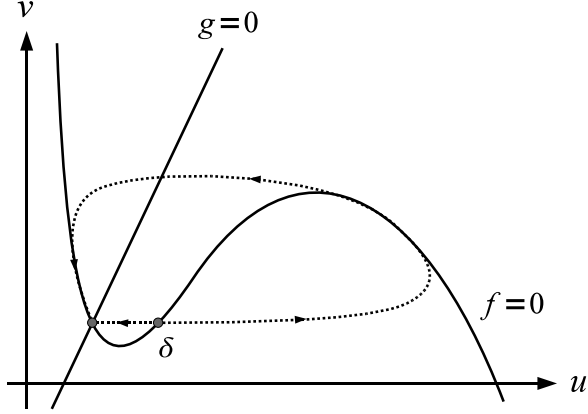}
\end{center}
\caption{The phase diagram for the excitable BZ reaction.}
\label{vf_bz}
\end{figure}

Let $f(u,v):=\frac{1}{\epsilon}[ u(1-u)-\frac{b v(u-a)}{u+a}]$ and $g(u,v):=u-v$.
Figure \ref{vf_bz} shows the phase diagram for the excitable state of the BZ reaction.
The null clines that are obtained from $f(u,v)=g(u,v)=0$ are shown by solid lines and a typical orbit for the excitable state is shown by a dashed line.  
The intersecting point of $f=0$ and $g=0$ is a stable point if there is no diffusion, 
however it becomes unstable when 
the strength of the perturbation (mainly due to the diffusion effects) exceeds a certain threshold $\delta$.
Then the state of the medium becomes unstable and changes along the dashed line shown in the phase diagram, until it returns to the equilibrium point once again.
The repetition of this process induces spacial patterns such as spiral waves.

\subsection{The CA Model}
Our CA model for Eq. (\ref{oregonator}) is described in the form of Eq. (\ref{camodel}), introduced in section 2.2.
According to Eq. (\ref{oregonator}), we put $\boldsymbol{u}^{t}_{mn}:=(u^{t}_{mn}, v^{t}_{mn})^{T}$, $\boldsymbol{G}(\boldsymbol{u}^{t}_{mn}):=(G_{u}(\boldsymbol{u}^{t}_{mn}), G_{v}(\boldsymbol{u}^{t}_{mn}))^{T}$ and $\boldsymbol{R}(\boldsymbol{\hat{u}}^{t}_{mn}):=(R_{u}(\hat{u}^{t}_{mn}), R_{v}(\hat{v}^{t}_{mn}))^{T}$.

\begin{figure}[h]
\begin{center}
\includegraphics[width=13cm]{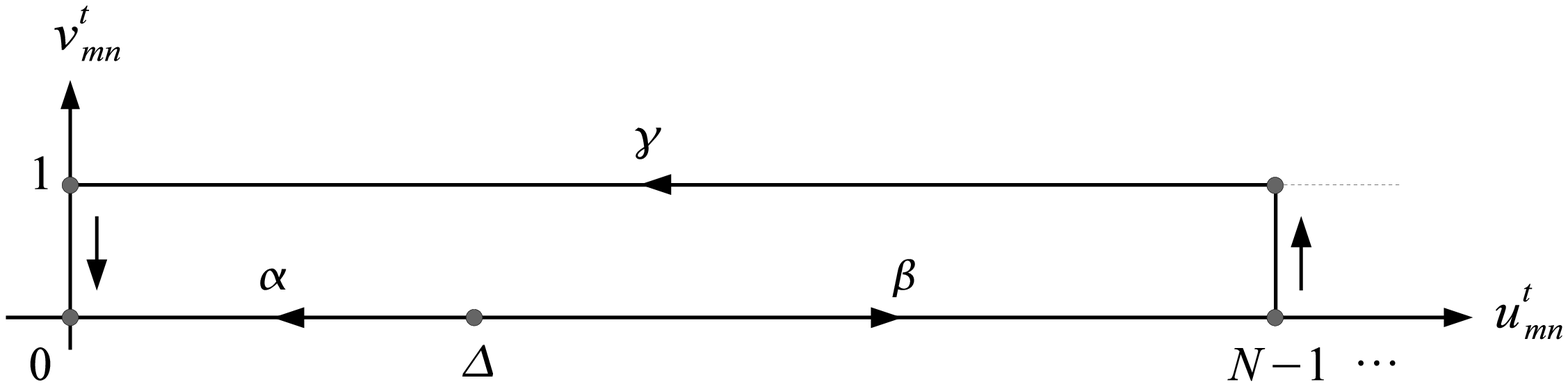}
\end{center}
\caption{Outline of the discrete vector field $\boldsymbol{G}(\boldsymbol{u}^{t}_{mn})$ in \tablename\ref{table1}.}
\label{dvf_bz}
\end{figure}

First, we define the discrete vector field $\boldsymbol{G}(\boldsymbol{u}^{t}_{mn})$ by imitating the solution orbit of the phase diagram shown in Figure \ref{vf_bz}.
The discrete vector field we configured is shown in Table \ref{table1} and illustrated in Figure \ref{dvf_bz}.
We can see that Figure \ref{dvf_bz} is a simplification of the phase diagram of Figure \ref{vf_bz}.
Let $u^{t}_{mn} \in \Bbb{Z_{+}}$, $v^{t}_{mn} \in \{0, 1\}$ in this example of an excitable BZ reaction.
Parameters $\alpha$, $\beta$ and $\gamma$ control the rate of reaction, and $\Delta$ is the threshold which determines whether the excitation occurs or not.
Parameter $N \in \Bbb{Z_{+}}$ denotes the expected maximum value of variable $u^{t}_{mn}$.
All of these five parameters are positive integers.

The value of the discrete vector field $\boldsymbol{G}(u^{t}_{mn}, v^{t}_{mn})$ is determined in function of the values of $u^{t}_{mn}$ and $v^{t}_{mn}$, as shown in Table \ref{table1}.
For $0 \leq u^{t}_{mn} < \Delta, v^{t}_{mn} = 0$, the state of the medium returns to the equilibrium point with velocity $\alpha$, because the state cannot exceed the threshold $\Delta$ due to inadequate diffusion effects.
For $\Delta \leq u^{t}_{mn} < N-1-\beta, v^{t}_{mn} = 0$, the variable $u^{t}_{mn}$ increases at the rate $\beta$ until $u^{t}_{mn} \geq N-1-\beta$.
In the range $\gamma < u^{t}_{mn}, v^{t}_{mn} = 1$,  the variable $u^{t}_{mn}$ decreases at the rate of $\gamma$.
In the two remaining ranges, the variable $v^{t}_{mn}$ changes between 0 and 1.
Here we assumed $v^{t}_{mn} \in \{0, 1\}$, because two states are sufficient to separate the phases, according to the medium, corresponding to the variable $v^{t}_{mn}$.

\begin{table}[h]
\caption{Example of discrete vector field $\boldsymbol{G}(\boldsymbol{u}^{t}_{mn})$ for an excitable medium.}
\begin{tabular}{c|c}
$(u^{t}_{mn}, v^{t}_{mn}) = \boldsymbol{u}^{t}_{mn}$ & $(G_{u}(\boldsymbol{u}^{t}_{mn}), G_{v}(\boldsymbol{u}^{t}_{mn})) = \boldsymbol{G}(\boldsymbol{u}^{t}_{mn})$\\
\noalign{\hrule height1pt}
$(0 \leq u^{t}_{mn} < \Delta, v^{t}_{mn} = 0)$ & $(\rm{max}[u^{t}_{mn} - \alpha, 0], 0)$\\
\hline
$(\Delta \leq u^{t}_{mn} < N-1-\beta, v^{t}_{mn} = 0)$ & $(u^{t}_{mn} + \beta, 0)$\\
\hline
$(N-1-\beta \leq u^{t}_{mn}, v^{t}_{mn} = 0)$ & $(N-1, 1)$\\
\hline
$(\gamma < u^{t}_{mn}, v^{t}_{mn} = 1)$ & $(u^{t}_{mn} - \gamma, 1)$\\
\hline
$(0 \leq u^{t}_{mn} \leq \gamma, v^{t}_{mn} = 1)$ & $(0, 0)$
\end{tabular}
\label{table1}
\end{table}

In Eq. (\ref{oregonator}), the excitation of excitable media is determined by the diffusion of $u$, i.e., $d_u \gg d_v$.
Hence, in this example, we only consider the random walk of the variable $u^{t}_{mn}$
and we simulate the time evolution by putting $P={\rm{diag}}(p_{u}, 0)$.

Consequently, our CA model for the excitable BZ reaction is given as:
\begin{equation}
\begin{split}
\label{camodel_bz}
& u^{t+1/2}_{mn} = u^{t}_{mn} +   R_{u}(\hat{u}^{t}_{mn}),\\
& v^{t+1/2}_{mn} = v^{t}_{mn},\\
& u^{t+1}_{mn} = G_{u}(\boldsymbol{u}^{t+1/2}_{mn}),\\
& v^{t+1}_{mn} = G_{v}(\boldsymbol{u}^{t+1/2}_{mn}).
\end{split}
\end{equation}

\subsection{Numerical Results}
In this section, we show several time evolution patterns obtained from Eq. (\ref{camodel_bz}) and discuss the results.

\subsubsection{Patterns}
\begin{figure}[h]
\begin{center}
\includegraphics[width=5cm]{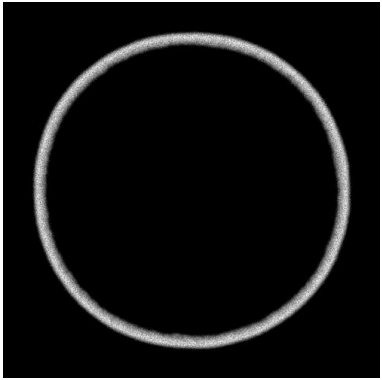}
\end{center}
\caption{Single trigger wave. $500 \times 500$ cells, $t =5850 $, $N =100$, $p_{u} = 0.2$, $\Delta=21$, $\alpha=1$, $\beta=1$, $\gamma=1$.}
\label{ring}
\end{figure}

The first example is a single trigger wave (Figure \ref{ring}). 
It is produced by the initial condition $u^{0}_{mn} = [h \cdot {\rm exp}(-((m-L/2)^{2}+(n-L/2)^{2})/w^{2})]$ and $v^{0}_{mn} = 0$ on a 2-dimensional square area with $L \times L$ cells.
Here $[\hspace{1mm}]$ is Gauss' symbol, i.e.: $[x]$ is the largest integer that is less than or equal to $x$ and $h$ and $w$ are positive real numbers.
We observed that from a pulse triggered at the center, a ring-shaped wave spreads outwards in an almost isotropic fashion.

\begin{figure}[h]
\begin{center}
\includegraphics[width=5cm]{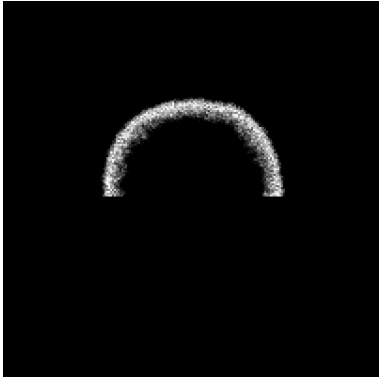}
\end{center}
\caption{A cut trigger wave. $200 \times 200$ cells, $t =250 $, $N = 30$, $p_{u} = 0.2$, $\Delta=6$, $\alpha=1$, $\beta=2$, $\gamma=1$.}
\label{cutring}
\end{figure}

\begin{figure}[h]
\begin{center}
\includegraphics[width=5cm]{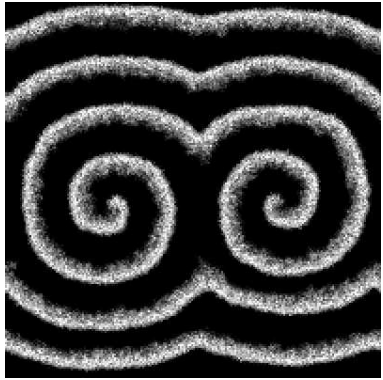}
\end{center}
\caption{Spiral wave. $200 \times 200$ cells, $t =963 $, $N = 30$, $p_{u} = 0.2$, $\Delta=6$, $\alpha=1$, $\beta=2$, $\gamma=1$.}
\label{spiral}
\end{figure}

In order to generate a spiral wave, appropriate initial conditions are necessary. Firstly, we generate a single trigger wave like the one shown in Figure \ref{ring}.
Then we cut off one part of the ring pattern as shown in Figure \ref{cutring} and use the remainder as the initial state.
The spiral wave obtained from our model is shown in Figure \ref{spiral}.

\begin{figure}[h]
\begin{center}
\includegraphics[width=5cm]{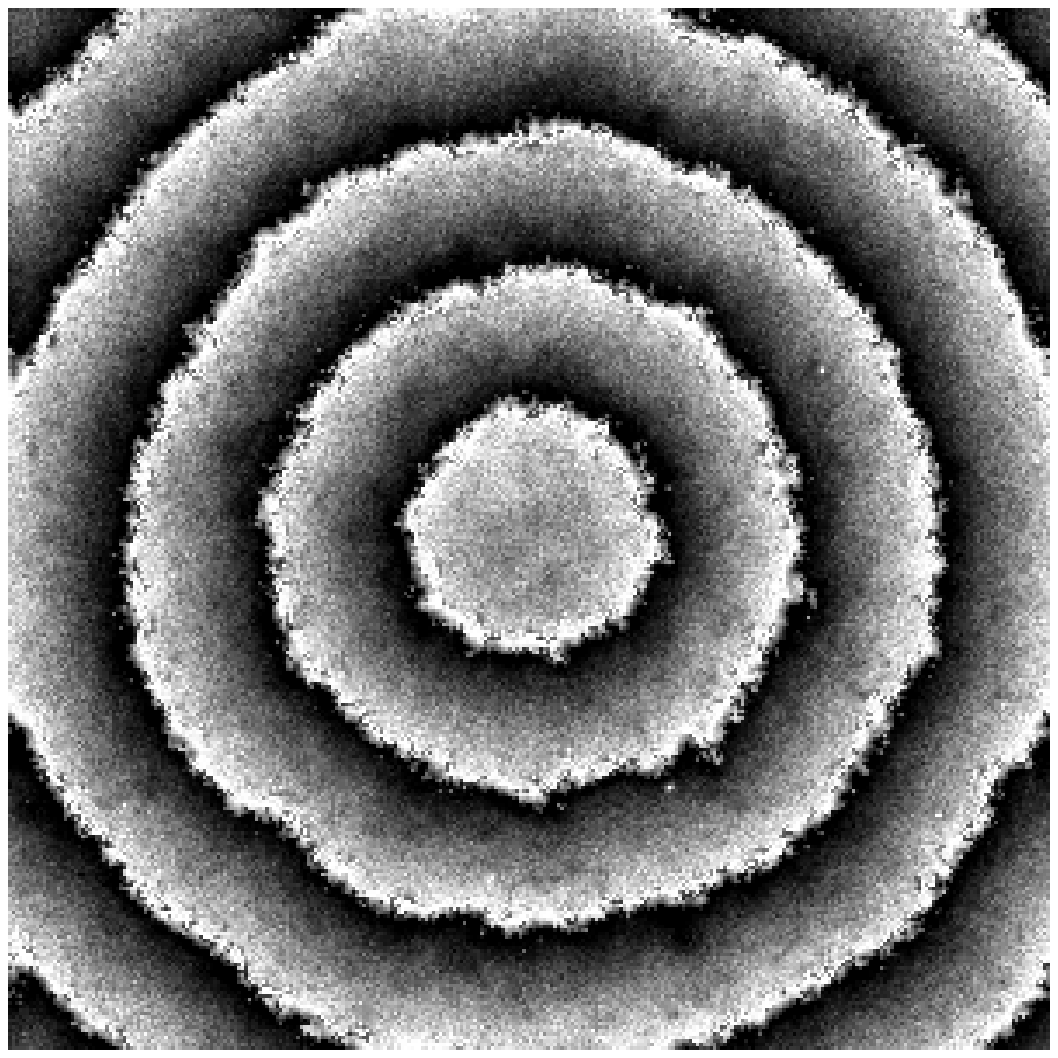}
\end{center}
\caption{Target pattern. $300 \times 300$ cells, $t =626 $, $N = 100$, $p_{u} = 0.04$, $\Delta=2$, $\alpha=1$, $\beta=10$, $\gamma=1$.}
\label{target}
\end{figure}

The third example is a target pattern (Figure \ref{target}).
The initial condition is the same as in the case of the trigger wave, but with different parameters.
From the central pulse, ring-shaped waves are produced repeatedly.

\subsubsection{Anisotropy}
\begin{figure}[h]
\begin{center}
\includegraphics[width=14cm]{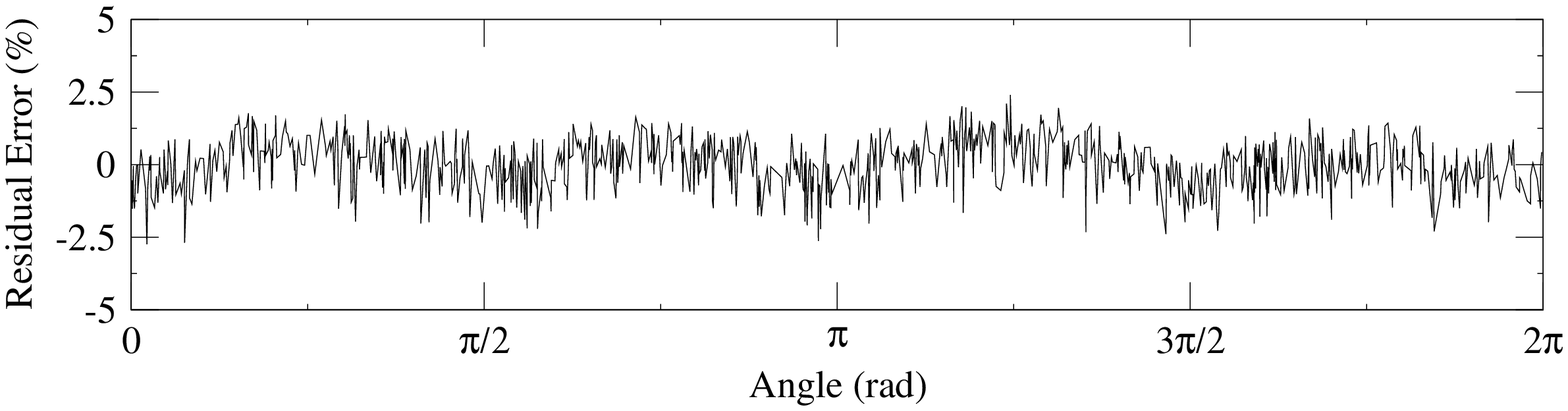}
\end{center}
\caption{Plot of the anisotropy of the model, measured as the residual error of the trigger pattern in \figurename\ref{ring}.}
\label{residualerror}
\end{figure}

\begin{figure}[h]
\begin{center}
\includegraphics[width=14cm]{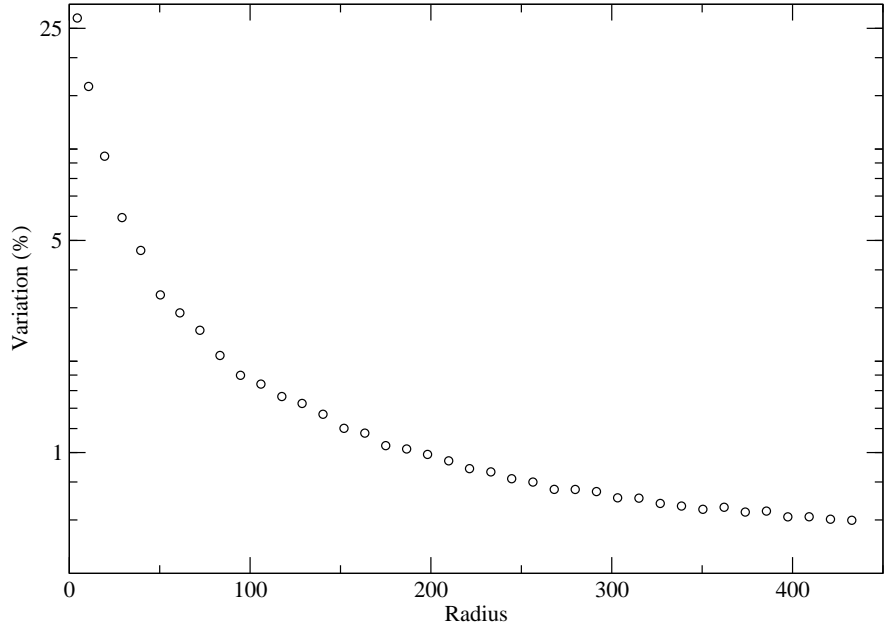}
\end{center}
\caption{The relationship between the radius of the ring and the variation of the radius.}
\label{variation}
\end{figure}

We evaluate the anisotropy of the trigger wave in Figure \ref{ring} by measuring the residual error when compared with the average radius of the ring, and plot it in Figure \ref{residualerror} as a function of the propagation direction. 
We find that the wave fronts of the trigger wave propagate in each direction with an anisotropy in the range of $\pm 2.5$ percent.
In Figure \ref{variation} we plot the variation from the circle as a function of the radius of the trigger wave.
It shows that the trigger wave indeed grows closer to a complete circle as the radius (and therefore time) increases.

\begin{figure}[h]
\begin{center}
\includegraphics[width=14cm]{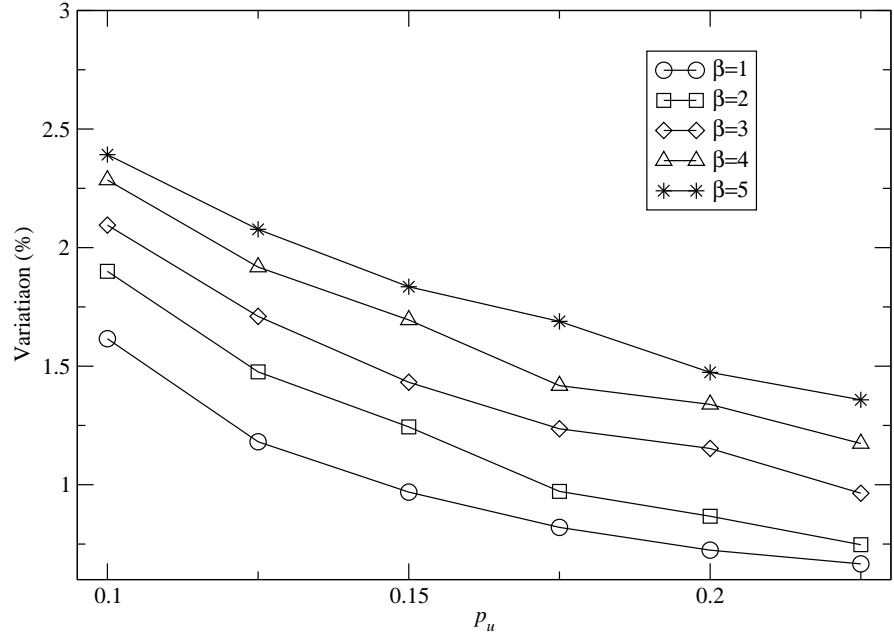}
\end{center}
\caption{The parametric dependency of the anisotropy of the model for the parameter $\beta$. $N=100$, $\Delta=15$, $\alpha=1$, $\gamma=1$.}
\label{delta=15}
\end{figure}

\begin{figure}[h]
\begin{center}
\includegraphics[width=14cm]{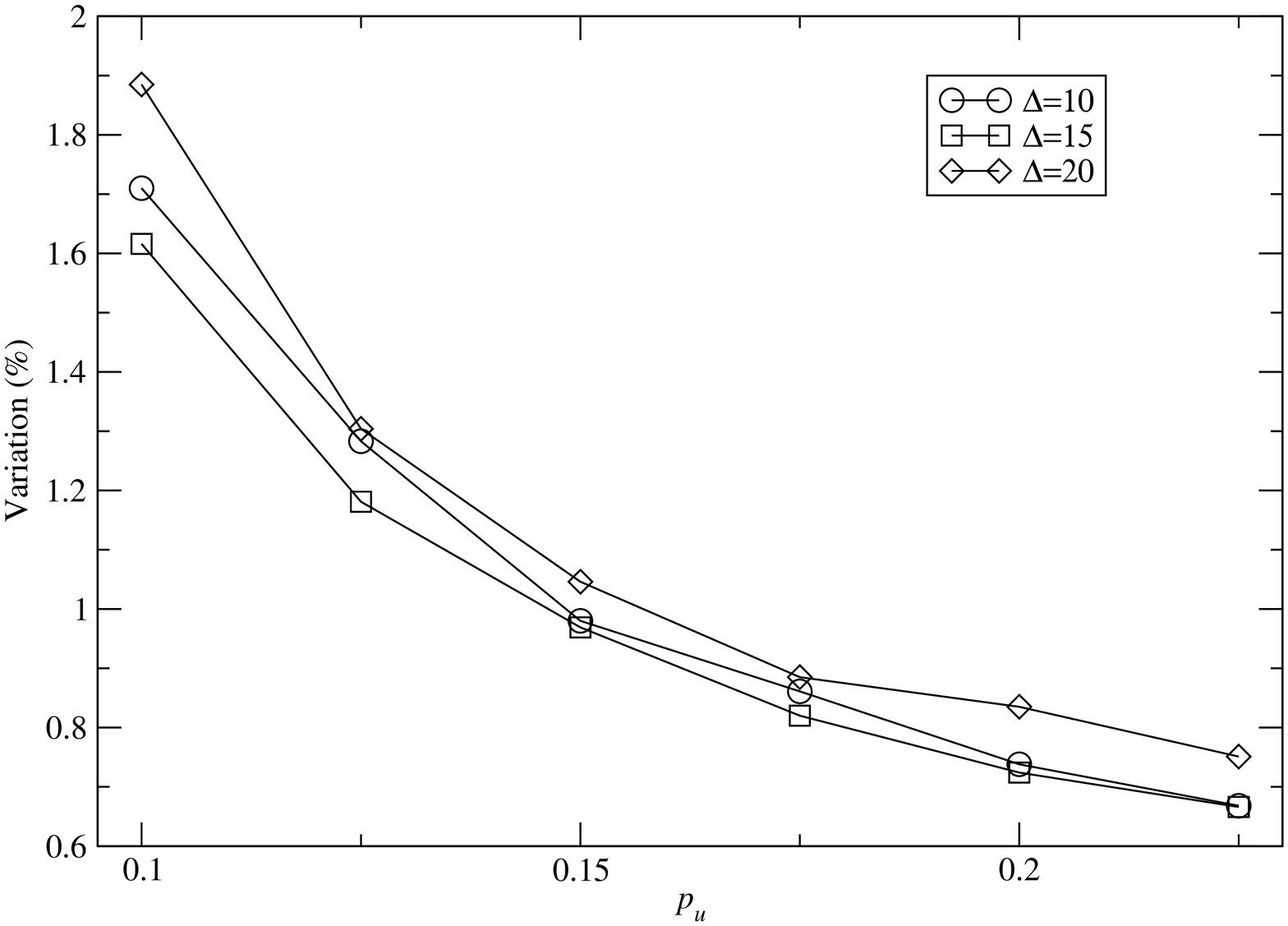}
\end{center}
\caption{The parameter dependence of the anisotropy of the model on the parameter $\Delta$. $N=100$, $\alpha=1$, $\beta=1$, $\gamma=1$.}
\label{beta=1}
\end{figure}

Next, we evaluate the parameter dependency of the anisotropy for the patterns observed from our model.
In Figures \ref{delta=15} and \ref{beta=1}, the variations of the wavefronts of trigger waves propagated to radius 450, are plotted as a function of the transition probability $p_{u}$ of particles by changing parameters $\beta$ (Fig. \ref{delta=15}) and $\Delta$ (Fig. \ref{beta=1}).
We find that the patterns become more isotropic when diffusion becomes stronger, and that they tend to be isotropic for smaller $\beta$, but depend less on $\Delta$.

\subsubsection{Parameters}
\begin{figure}[h]
\begin{center}
\includegraphics[width=14cm]{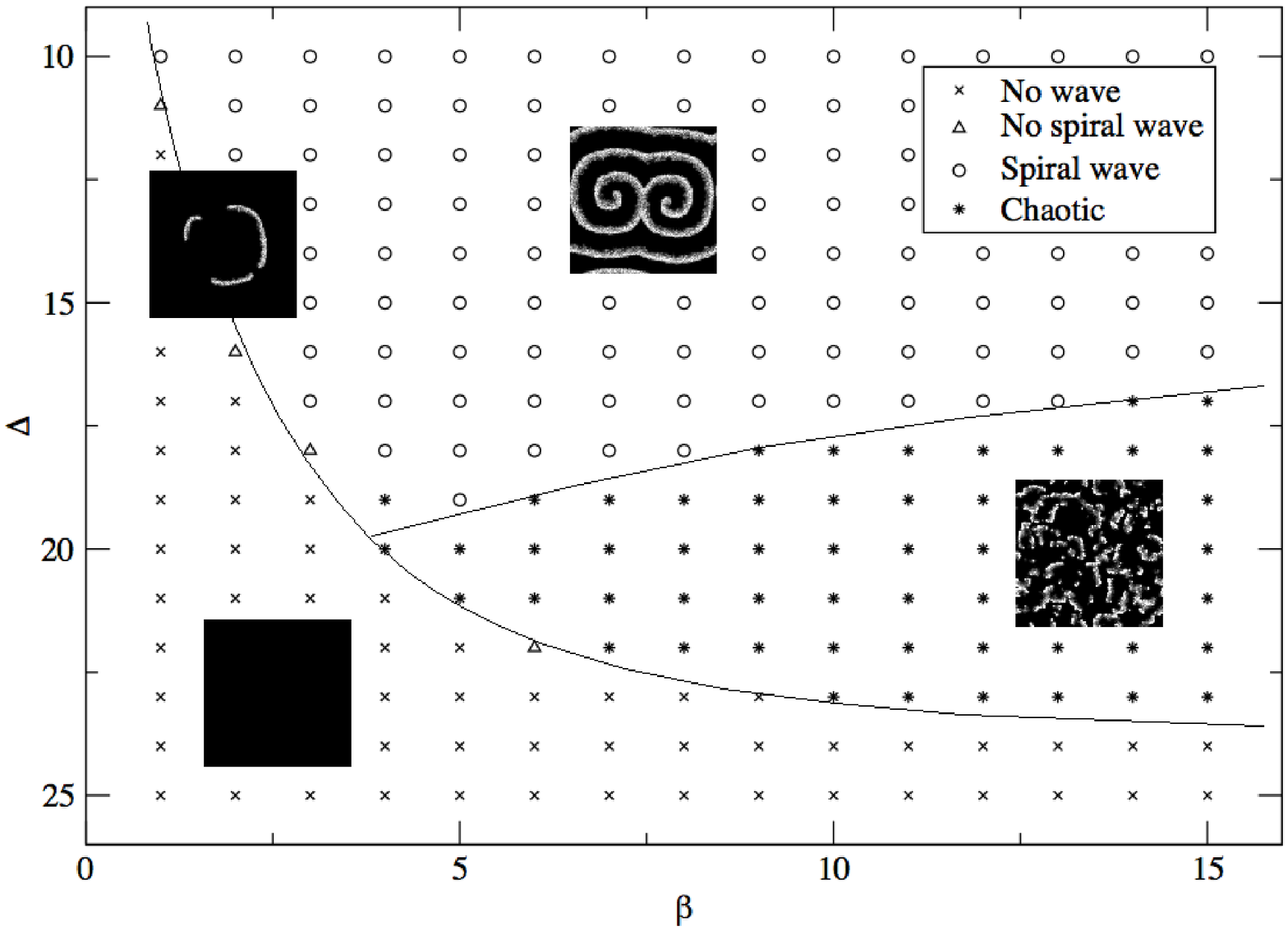}
\end{center}
\caption{The phase diagram of the patterns obtained from our CA.
$N = 50$, $p_{u} = 0.1$, $\alpha=\gamma=1$.}
\label{patternphase_p=0.1}
\end{figure}

Our CA model has six parameters $N$, $p_u$, $\Delta$, $\alpha$, $\beta$ and $\gamma$.
The parameter $\Delta$ corresponds to $b$ in  Eq. (\ref{oregonator}) and defines the threshold of excitation; $\beta$ corresponds to $1/\epsilon$ which defines the  excitability of reaction.
Figure \ref{patternphase_p=0.1} is the phase diagram of our model obtained by changing the values of $\Delta$ and $\beta$.
The remaining parameters are chosen as $N = 50$, $p_u=0.1$,$\alpha=\gamma=1$.
We confirm that spiral waves can be generated in a wide region of the parameters.
In an even larger parameter range, we find there exist three different regions; one without propagating waves, one allowing for trigger waves (or broken trigger waves), and one chaotic pattern region.

The parameter range for the spiral waves obtained from Eq. (\ref{oregonator}) has been examined in detail by Jahnke and Winfree \cite{jahnke}.
The general tendency is that, when $b$ is sufficiently large, the spiral wave does not appear and for $1/\epsilon \gg 1$ the spiral wave destabilizes and chaotic behaviour is observed.
As shown in Figure \ref{patternphase_p=0.1}, the phase diagram of our model has the same feature as that for Eq. (\ref{oregonator}), and we may conclude that the parameters $\Delta$ and $\beta$ play the same role as the control parameters of the reaction terms in the reaction-diffusion equation.

\section{summary}
We have proposed a method to construct a CA model corresponding to a reaction-diffusion equation, in which the diffusion effect is replaced by a random walk with transition probability matrix $P$ and the reaction by discrete vector fields. 
The model can include control parameters for both diffusion and reaction, as is the case in the reaction-diffusion equation.
As an example, we have shown that our model can successfully reproduce the patterns of BZ reaction.
Applications to other reaction-diffusion systems such as FitzHugh-Nagumo equation \cite{nagumo} are interesting future problems.
In the present method, however, there are still various candidates for the time evolution rules, depending on the choice of the discrete vector fields that are supposed to have similar features to those of the continuous vector fields, given by the reaction-diffusion equation.
Of course we can also adopt discrete vector fields by investigating the reaction process from a microscopic point of view.
The determination a suitable time evolution rule will depend on the individual phenomenon and we will investigate this problem extensively in the future. 

\section*{Acknowledgments}
We would like to thank professors Hiroshi Tanaka and Ralph Willox for useful discussions and comments.

\end{document}